\documentclass[reprint,superscriptaddress,aps,prd,floatfix,nofootinbib,bibnotes]{revtex4-1}
\usepackage[colorlinks=true,linkcolor=black,citecolor=blue, urlcolor=blue,bookmarks=false]{hyperref}
\hypersetup{breaklinks=true}
\usepackage[utf8]{inputenc}
\usepackage{natbib}
\usepackage{bm,mathtools,cleveref}
\usepackage{array}
\usepackage{epsfig}
\usepackage{amsmath}
\usepackage{amssymb}
\usepackage{multirow}
\usepackage{graphicx,subcaption}
\usepackage{comment}
\usepackage{enumitem}
\usepackage[normalem]{ulem}  % \sout{old text} for strikeout
\usepackage[dvips]{color} % For blue in-text comments and additions

\renewcommand{\sout}{\bgroup \color{red} \ULdepth=-.5ex \ULset}

\newcommand{\vev}[1]{\langle{#1}\rangle}

\newcommand{\qv}{\vec{q}\,}

\begin{document}
\title{Heavy quarkonium with finite three momentum near $T_c$}

\author{HyungJoo Kim}
\email{hugokm0322@gmail.com}
\affiliation{Department of Physics and Institute of Physics and Applied Physics, Yonsei University, Seoul 03722, Korea}
\author{Seokwoo Yeo}
\email{ysw351117@gmail.com}
\affiliation{Department of Physics and Institute of Physics and Applied Physics, Yonsei University, Seoul 03722, Korea}
\author{Sungtae Cho}
\email{sungtae.cho@kangwon.ac.kr}
\affiliation{Division of Science Education, Kangwon National University, Chuncheon 24341, Korea}
\author{Su Houng Lee}%
\email{suhoung@yonsei.ac.kr}
\affiliation{Department of Physics and Institute of Physics and Applied Physics, Yonsei University, Seoul 03722, Korea}

\date{\today}
\begin{abstract}
We investigate the non-trivial 3-momentum effects on the masses of heavy quarkonium states that are moving in a hot medium using QCD sum rules.
For all charmonium states, we observe a negative mass shift near $T_c$ that is less than 3$\%$ at a momentum of 1$\rm{GeV}$.
Specifically, we first investigate the difference between the longitudinal and transverse modes of both $J/\psi$ and $\chi_{c1}$. We find that the transverse mode of the $J/\psi$ experiences larger modification than the longitudinal mode, while the $\chi_{c1}$ has the opposite behavior. 
By comparing the $\eta_c$ and $\chi_{c0}$, and also the unpolarized $J/\psi$ and $\chi_{c1}$, we recognize that the P-wave particles have stronger momentum dependencies on their masses than the S-wave ones.
 We also find $\Upsilon$(1S) has negligible 3-momentum dependence compared to the charmonium states, e.g. less than 0.01$\%$ even at 1.4$T_c$ and at a momentum of 4$\rm{GeV}$.
\end{abstract}

\maketitle
\section{Introduction}
The study of quarkonium in a thermal medium is of great importance in our understanding of the physics of strongly-interacting matter that existed in the early universe. In particular, heavy quarkonium in a hot medium has long been of interest ever since the seminal work by Matsui and Satz\cite{Matsui:1986dk} that suppression of quarkonium yields in heavy-ion collisions could be a signal for the formation of a deconfined quark-gluon plasma. As temperature increases, quarkonium states undergo spectral modification, such as mass shift and broadening, and eventually will melt and merge with the continuum. Over the past decade, many theoretical approaches, such as lattice simulations, effective field theories, and spectral reconstruction, are developed and employed to understand the properties of quarkonium in a hot medium. Moreover, the recent recognition that the in-medium interquark potential is complex-valued\cite{Laine:2006ns} stressed the necessity of dynamical description for quarkonium melting and led to a new framework of the open quantum system. See Refs.\cite{Brambilla:2010cs,Rothkopf:2019ipj} for comprehensive reviews.

Meanwhile, most theoretical studies assume quarkonium is at rest when computing its in-medium properties. Because Lorentz symmetry is broken in the presence of the medium, finite 3-momentum makes any computation more complicated and thus often omitted. However, most particles produced in real experiments propagate in a medium with non-zero momentum. For a more realistic analysis, therefore, it is  necessary to take into account non-trivial effects coming from finite 3-momentum. In general, it is expected that two types of effects appear for a particle moving in a medium. First, a particle will not follow the standard energy-momentum dispersion relation, i.e. $E^2-\qv^2=m^2$, where $E$, $\qv$ and $m$ denote the energy, 3-momentum and invariant mass of a particle, respectively. Instead, it will follow a modified version, i.e. $E^2-\qv^2=m(\qv)^2$, which indicates that the mass of a particle moving in a medium is not a invariant quantity any more  but possibly depends on the 3-momentum. 
Second, if a particle has spin, its different polarization states that have the same dispersion relation in a vacuum may behave differently. For example, the transverse and longitudinal modes of massive spin-1 particles will follow separate modified dispersion relations in a medium. In other words, by tracking the 3-momentum dependence on the mass spectrum we might be able to distinguish the  polarization states of particles\cite{Park:2022ayr}.
Consequently, the overall mass shift can vary depending on the polarization states as well as the size of 3-momentum.
Therefore, when we read off the mass shift from the spectral change of quarkonium in a medium, the finite momentum effect should be identified as distinct from the pure mass shift typically defined at zero momentum.
%%%%3%%%%%%%%
%Since the direction and magnitude of the mass shift can vary depending on the polarization, this effect is particularly important when we read off mass shift from the spectral change of quarkonium in a medium. Therefore, finite momentum effects should be identified as distinct from the pure mass shift typically defined at zero momentum.

There are some lattice simulations that have studied non-trivial momentum effects on the dispersion relation of heavy quarkonium at finite temperature\cite{Oktay:2010tf,Ding:2012pt,Aarts:2012ka,Ikeda:2016czj,Ding:2017eem}. All of them are based on the maximum entropy method(MEM) to reconstruct the spectral function from lattice simulation for the euclidean correlator. The $\eta_c$ and $J/\psi$ were studied in Refs.\cite{Oktay:2010tf,Ding:2012pt,Ikeda:2016czj}. The $\Upsilon$(1S) and $\eta_b$ are studied in Ref.\cite{Aarts:2012ka} using lattice NRQCD. In Ref.\cite{Ding:2017eem}, S-wave states for both charmonium and bottomonium are studied. In spite of these many efforts, however, none of them made any definite conclusion on the non-trivial effects because of the significant size of uncertainties.

In this work, we investigated the non-trivial 3-momentum effects on the masses of heavy quarkonium states that are moving in a hot medium using QCD sum rules. 
This method has already been applied to study light vector mesons, such as $\rho$, $\omega$, and $\phi$, that are moving in a dense medium\cite{Lee:1997zta,Leupold:1998bt,Kim:2019ybi}, so we basically follow the same strategy discussed there. Specifically, we included the Wilson coefficients which are responsible for the non-trivial 3-momentum dependence to the typical QCD sum rules for the quarkonium at rest.
We first studied the difference between the longitudinal and transverse modes of $J/\psi$ and $\chi_{c1}$. To contrast the difference between S-wave and P-wave particles, we then compared the $\eta_c$ and $\chi_{c0}$, and also the unpolarized $J/\psi$ and $\chi_{c1}$. We also studied $\Upsilon$(1S) as a representative of bottomonium states while distinguishing the longitudinal and transverse modes. We hope that our work will provide a guideline on how much uncertainty must be reduced in future lattice simulations to reveal the genuine non-trivial effects on the dispersion relation of heavy quarkonium at finite temperature.

This paper is organized as follows. In Section II, we give a brief description of the formalism of QCD sum rules for a particle moving in a medium.
Section III is devoted to the detailed results obtained in this study. 
The paper is summarized and concluded in Section IV. In the Appendix, we display the explicit forms of  Wilson coefficients which are responsible for the non-trivial 3-momentum dependence.

\section{QCD sum rules with finite momentum}

To study heavy quarkonium states in the pseudoscalar(P), scalar(S), vector(V), and axialvecotr(A) channels, we consider a two-point correlator,
\begin{align}
    \Pi^J(\omega,\qv)=i\int d^4x e^{iq\cdot x}\vev{T\{j^J(x) j^J(0) \}},\label{correlationftn}
\end{align}
where the superscript $J$ indicates each channel that has the following current structure: $j^{P}=i\bar{h}\gamma_{5}h$, $j^{S}=\bar{h}h$, $j_{\mu}^{V}=\bar{h}\gamma_{\mu}h$, and $j_{\mu}^{A}=(q_{\mu}q_{\nu}/q^{2}-g_{\mu \nu})\bar{h}\gamma_{5}\gamma^\nu h$. Here, $h$ represents the heavy quark field that can be either a $c$ or $b$ quark. 
For convenience, we define the following dimensionless functions,
\begin{align}
    \widetilde{\Pi}{^{P,S}(\omega^{2},\qv^{2})}&=\frac{1}{q^{2}}\Pi^{P,S},\\
      \widetilde{\Pi}^{V,A}_{L}(\omega^{2},\qv^{2})&=\frac{1}{\qv^{2}}\Pi^{V,A}_{00},\\
    \widetilde{\Pi}^{V,A}_{T}(\omega^{2},\qv^{2})&=-\frac{1}{2}(\frac{1}{\qv^{2}}\Pi_{\mu}^{\mu,V,A}+\widetilde{\Pi}^{V,A}_{L}),\label{dimensionless correlator}
\end{align}
where the V and A channels are decomposed into the longitudinal($L$) and transverse($T$) modes.

In the deep Euclidean region where $\omega^2 \ll \infty$ with finite $|\vec{q\,}|$, one can compute these functions using the operator product expansion(OPE). As stressed in Ref.\cite{Kim:2019ybi}, it is more convenient to express the OPE results by changing variables from $(\omega^2,\vec{q\,}^2)$ to $(Q^2,\vec{q\,}^2)$ where $Q^2\equiv -\omega^2+\vec{q\,}^2$. 
After this substitution, it becomes transparent that $\vec{q\,}^2$ absorbed in $Q^2$ does not violate Lorentz symmetry while the remaining $\vec{q\,}^2$, which only appears in the Wilson coefficients of non-scalar operators, is responsible for the non-trivial 3-momentum effects on the dispersion relation.
In this work, OPE is considered up to dimension 4 gluon operators, i.e. scalar($G_0$) and twist-2($G_2$) gluon condensates, which are defined as
\begin{align}
     \vev{\frac{\alpha_{s}}{\pi}G_{\mu \alpha}^{a}G_{\nu}^{a,\alpha}}=\frac{1}{4}g_{\mu\nu}G_0+(u_{\mu}u_{\nu}-\frac{1}{4}g_{\mu \nu})G_{2},\label{G0 and G2}
\end{align}
where the medium four-velocity, denoted as $u_\mu$, is taken to be at rest. Because Wilson coefficients of these operators are already given in a covariant form for all relevant channels \cite{Reinders:1981si,Reinders:1984sr,Klingl:1998sr,Song:2008bd}, we can simply extract 3-momentum dependent terms after the change of variables. 

In the conventional analysis, we typically apply the Borel transform to the correlator with respect to $Q^2$,
\begin{align}
    \mathcal{M}^J (M^{2},\vec{q\,}^2)=\lim_{\mathclap{\substack{n,Q^{2} \to \infty,\\{Q^{2}/n=M^{2}}}}}\frac{(Q^{2})^{n+1}}{n!}\left(-\partial_{Q^2}\right)^{n}\widetilde{\Pi}^J(Q^2,\vec{q\,}^2),\label{Borel transform}
\end{align}
where $\mathcal{M}^J (M^{2},\vec{q\,}^2)$ indicates the Borel transformed correlator with $M$ being the Borel mass parameter. After this transform, the final expression of OPE result can be written as
\begin{align}
    \mathcal{M}^J_{\text{OPE}}(M^2,\vec{q\,}^2)=&e^{-\nu} A^J(\nu)\Big[1+\alpha_{s}a^J(\nu)+b^J(\nu)\phi_{b}\nonumber\\
    &+\Big\lbrace c^J(\nu)+\frac{\vec{q\,}^2}{m_h^2}d^J(\nu)\Big\rbrace\phi_{c}\Big],\label{borelOPE}
\end{align}
where $\nu=4m_h^2/M^2$, $\phi_b=\frac{4\pi^2}{9(4m_h^2)^2}G_0$, and $\phi_c=\frac{4\pi^2}{3(4m_h^2)^2}G_2$. For input parameters, we use $m_c(p^2=-m_c^2)=1.262\,\rm{GeV}$, $\alpha_s(8m_c^2)=0.21$ for charmonium states and  $m_b(p^2=-m_b^2)=4.12\,\rm{GeV}$, $\alpha_s(8m_b^2)=0.158$ for bottomonium states\cite{Morita:2009qk}. The values of $G_0$ and $G_2$ at finite temperature are taken from pure SU(3) lattice gauge theory\cite{Boyd:1996bx,Morita:2009qk}. The Wilson coefficients, $A^J(\nu)$, $a^J(\nu)$, $b^J(\nu)$, and $c^J(\nu)$, have  no distinction depending on the polarization states and their explicit forms can be found in Refs.\cite{Bertlmann:1981he,Morita:2009qk}.  In this work, we derive $d^J(\nu)$ for the first time and display their explicit forms in the appendix. The difference in the functional form between $d^{V,A}_L(\nu)$ and $d^{V,A}_T(\nu)$ explicitly show why the longitudinal and transverse modes of spin-1 quarkonium should have different behaviors in a medium.

By using analyticity of the correlator, one can connect Eq.\eqref{borelOPE} to the integral of the spectral function,
\begin{align}
    \mathcal{M}^J_{OPE}(M^2,\qv^2)=\int^\infty_{-\qv^2}ds e^{-s/M^2} \rho^J(s,\qv^2).\label{boreldispersion}
\end{align}
The spectral function, $\rho^J(s,\qv^2)$, is often modeled to have a single ground state pole and perturbative continuum,
\begin{align}
    \rho^J(s,\qv^2)\approx &f(\vec{q\,}^2)\delta(s-m_g^2(\qv^2))\nonumber\\
    &+\frac{1}{\pi}\text{Im}\widetilde{\Pi}^{J,\text{pert}}(s)\theta(s-s_0(\qv^2)),\label{spectralftn}
\end{align}
where all non-trivial 3-momentum dependence is assumed to be involved in the three spectral parameters, $f(\vec{q\,}^2)$, $m_g(\vec{q\,}^2)$, and $s_0(\vec{q\,}^2)$, which denote the residue, ground state mass, and threshold, respectively\cite{Kim:2019ybi}. The explicit forms of  $\text{Im}\widetilde{\Pi}^{J,\text{pert}}(s)$ can be found in Refs.\cite{Reinders:1981si,Morita:2009qk}. 

From Eq.(\ref{boreldispersion}) and the simple model of spectral function, the ground state mass for a given temperature and 3-momentum can be expressed as,
\begin{align}
    m_g(M^2,s_0)=\sqrt{-\frac{\frac{\partial}{\partial(1/M^2)}\bar{\mathcal{M}}^J(M^2,s_0)}{\bar{\mathcal{M}}^J(M^2,s_0)}},
\end{align}
where $\bar{\mathcal{M}}^J$ $(M^2,s_0)$ $=\mathcal{M}^J_{OPE}$ $(M^2,\qv^2)$ $-\int^\infty_{s_0}$ $ds$ $e^{-s/M^2}$ $\rho^J(s,\qv^2)$. This equation is reliable only inside a so-called Borel window, $(M_{min},M_{max})$, which is determined by competition between convergence of the OPE series and dominance of the pole contribution. Specifically, we assume the following two conditions:

\begin{align}
    M_{min} :\, & \Big|\alpha_{s}a^J(\nu)+b^J(\nu)\phi_{b}\nonumber\\
    &+\Big\lbrace c^J(\nu)+\frac{\vec{q\,}^2}{m_h^2}d^J(\nu)\Big\rbrace\phi_{c}\Big| \leq 0.3\\
    M_{max} :\, &\frac{\int^\infty_{s_0}ds e^{-s/M^2} \rho^J(s,\qv^2)}{\mathcal{M}^J_{OPE}(M^2,\qv^2)} \leq 0.4
\end{align}
Furthermore, we define the average value of the mass($\overline{m}_{g}$) and its uncertainty($\chi^{2}$) within the given Borel window as,
\begin{align}
    \overline{m}_{g}(s_{0})&=\int_{M_{min}}^{M_{max}}dM \frac{m_{g}(M^{2},s_{0})}{M_{max}-M_{min}}, \label{averaged mass}\\
    \chi^{2}(s_{0})&=\int_{M_{min}}^{M_{max}}dM \frac{(m_{g}(M^2,s_{0})-\overline{m}_{g}(s_{0}))^{2}}{M_{max}-M_{min}}.\label{chi square}
\end{align}
Finally, we find a threshold parameter that minimizes the uncertainty by scanning some range of threshold values. We then take the average mass at this threshold value as the final result for the ground state mass. This process is repeated while varying temperature, 3-momentum, channels, polarization states, and quark flavors.

\section{Results}
The ground state mass extracted from the above analysis depends on both the temperature($T$) and 3-momentum($\qv$). In the non-relativistic limit($|\qv|/ m_g\ll 1$), the energy-momentum dispersion relation can be expressed as,
\begin{align}
    E^2-\qv^2&=m_g^2(T,\qv^2)\nonumber\\
    &\approx m_g^2(T,0)-\alpha(T) \qv^2+\cdots,
\end{align}
where $m_g(T,0)$ denotes the rest mass at finite temperature and $\alpha(T)$ indicates the temperature dependent deviation parameter corresponding to the first-order coefficient of the non-trivial 3-momentum dependence on the ground state mass. Therefore, we are mainly interested in the signs and magnitudes of the deviation parameters for various quarkonium states that are moving at finite temperature.
To more intuitively illustrate how much the dispersion relation is modified in a medium, we also presented the energy ratio which is defined as
\begin{align}
    \frac{E(v)}{E(0)}\approx 1+\frac{1}{2}(1-\alpha(T) )v^2+\mathcal{O}(v^4), \label{energy ratio}
\end{align}
where $E(v)$ denotes the energy of a particle moving with finite velocity, $v=|\qv|/E(0)$, with $E(0)$ being equal to the rest mass $m_g(T,0)$. 
In this work, we studied the $J/\psi$, $\eta_c$, $\chi_{c0}$, and $\chi_{c1}$ of charmonium states, while only the $\Upsilon(1S)$  is studied as a representative of bottomonium states. 
For charmonium states, the 3-momentum is considered up to around 1$\rm{GeV}$, corresponding to $v^2\approx 0.1$. The maximum temperature of $J/\psi$ is 1.05$T_c$, while the others are considered up to 1.03$T_c$.
In the case of $\Upsilon(1S)$, the temperature is considered up to 1.4$T_c$ and the maximum 3-momentum is 4$\rm{GeV}$, corresponding to $v^2 \approx 0.18$.

\subsection{$J/\psi$ and $\chi_{c1}$ : Longitudinal vs Transverse}
\begin{figure}[h!]
\centering
\includegraphics[width=0.75\linewidth]{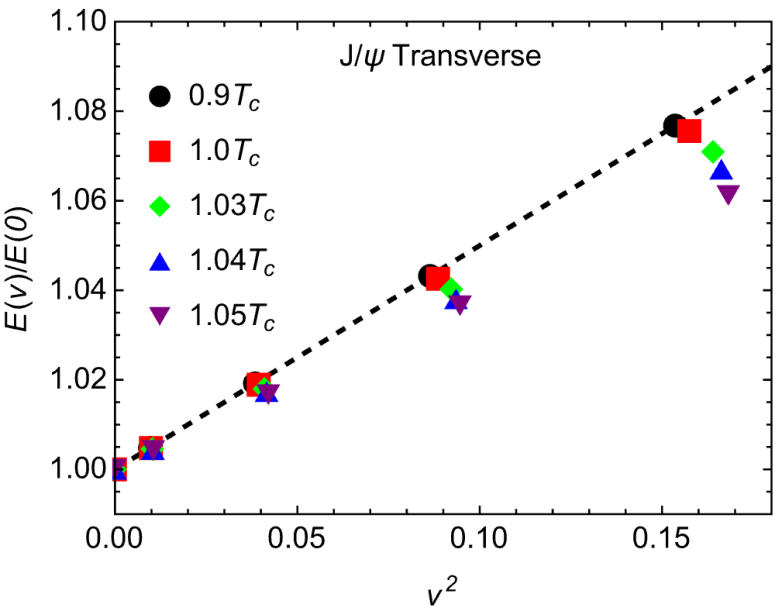}\\
\includegraphics[width=0.75\linewidth]{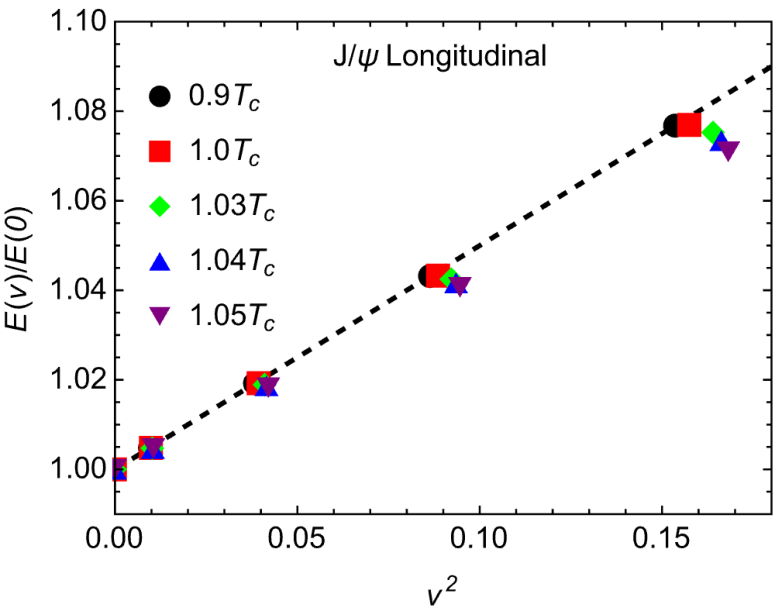}
\caption{Energy ratio of the transverse and longitudinal modes of $J/\psi$ at various temperatures and velocities. The dashed line shows $1+v^2/2$. }
\label{fig:jpsiratio}
\end{figure}

\begin{figure}[ht]
    \centering
    \includegraphics[width=0.75\linewidth]{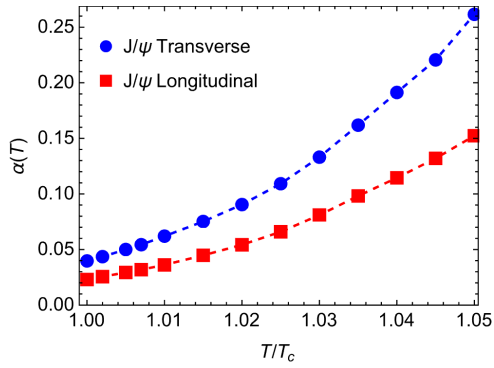}
    \caption{Deviation parameter $\alpha(T)$ of the $J/\psi$}
    \label{fig:jpsialpha}
\end{figure}

\begin{figure}[ht]
    \centering
    \includegraphics[width=0.75\linewidth]{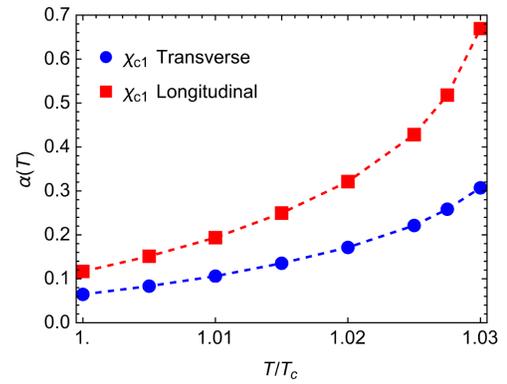}
    \caption{Deviation parameter $\alpha(T)$ of the $\chi_{c1}$}
    \label{fig:chic1alpha}
\end{figure}

Firstly, we examined the difference between the longitudinal and transverse modes of $J/\psi$. Their energy ratios are shown  as a function of $v^2$ in Fig.\ref{fig:jpsiratio}. 
As the black points$(0.9T_c)$ indicate the vacuum result, it is confirmed that both modes follow the trivial line, i.e. $1+v^2/2$, in the non-relativistic limit. But, they start to deviate from this line as the temperature increases. We found that the energies of both modes decrease with increasing momentum while the transverse mode deviates more than the longitudinal one. Similarly, it is also observed that the longitudinal mode of $J/\psi$ experiences smaller medium modification in Ref.\cite{Oktay:2010tf}, but the authors could not see a definite signal for deviation due to large uncertainties. Furthermore, we can also extract the deviation parameter  from fitting the energy ratio results with Eq.\eqref{energy ratio}. In Fig.\ref{fig:jpsialpha} we plot deviation parameters of both modes together as a function of temperature. It can be seen that the difference between the two polarization modes increases with increasing temperature. In addition, the deviation parameter of $\chi_{c1}$ was extracted in a similar manner as shown in Fig.\ref{fig:chic1alpha}. Unlike the $J/\psi$, however, we observed the opposite behavior for the $\chi_{c1}$, i.e. the transverse mode has the smaller deviation parameter than the longitudinal one.

\subsection{Charmonium : S-wave vs. P-wave}
\begin{figure}[h!]
    \centering
    \includegraphics[width=0.75\linewidth]{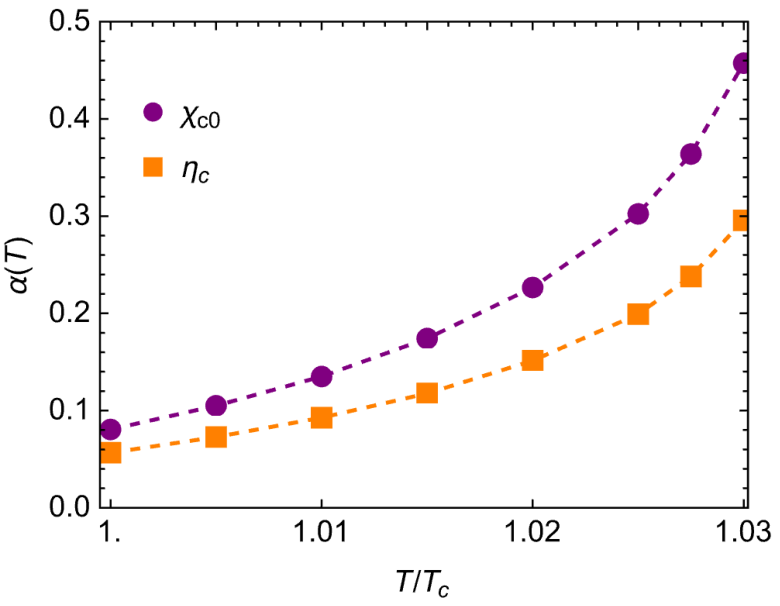}\\
    \includegraphics[width=0.75\linewidth]{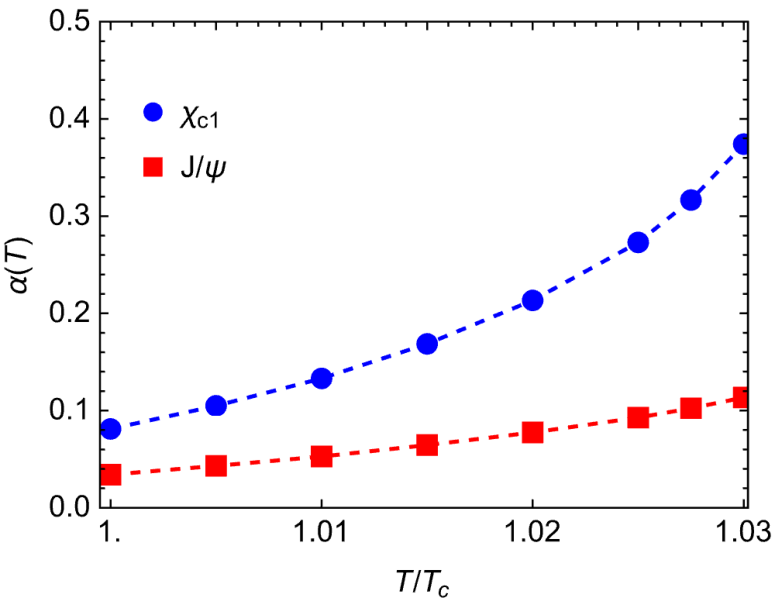}\\
    \caption{Deviation parameters compared between the $\eta_c$ and $\chi_{c0}$ (above), and between the unpolarized $J/\psi$ and $\chi_{c1}$ (below)}
    \label{fig:charmoniumalpha}
\end{figure}

\begin{figure}[h!]
    \centering
    \includegraphics[width=0.75\linewidth]{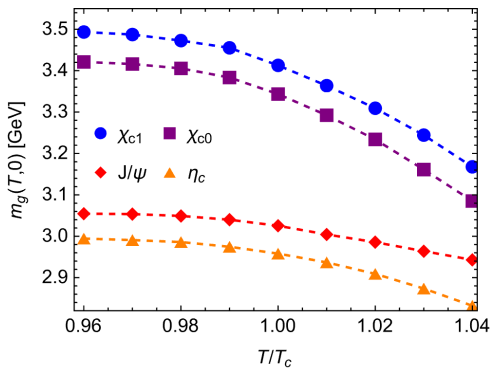}
    \caption{Temperature dependencies of rest masses $m_g(T,0)$ of the four charmonium states}
    \label{fig:restmass}
\end{figure}

For the next analysis, we compared the magnitude of deviation parameters of $\eta_c({}^1S_0)$ and $\chi_{c0}({}^3P_0)$, and also unpolarized $J/\psi({}^3S_1)$ and $\chi_{c1}({}^3P_1)$. The comparison results are shown together in Fig.\ref{fig:charmoniumalpha}. Overall, under the same total spin, it is apparent that the deviation parameters of P-wave particles$\,(\chi_{c0},\chi_{c1})$  are greater than those of S-wave particles$\,(\eta_c,J/\psi)$.
%Specifically, the magnitude of the deviation increases in the order of V, P, A, and S channels. 
This feature might be originated from the property that the P-wave particles experience stronger modification than the S-wave particles near $T_c$. The similar property can also be observed in the typical QCDSR analysis for the ground state masses at zero momentum as illustrated in Fig.\ref{fig:restmass}. When comparing rest masses of the four charmonium states near $T_c$, we also observed that the P-wave particles exhibit larger mass shifts than the S-wave ones. Therefore, these results suggest that as the rest mass is more susceptible to temperature changes, the dependence on 3-momentum will also increase.

\subsection{Bottomonium : $\Upsilon(1S)$}
\begin{figure}[h!]
    \centering
    \includegraphics[width=0.75\linewidth]{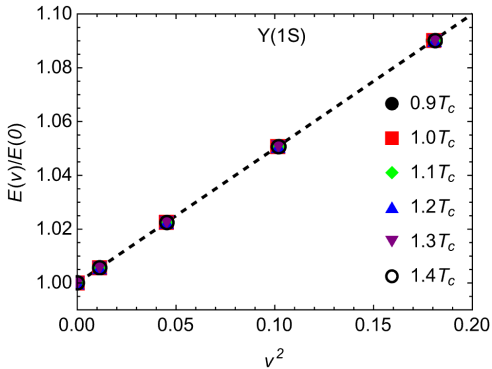}
    \caption{Energy ratio of the unpolarized $\Upsilon(1S)$ at various temperatures and velocities. The dashed line shows $1+v^2/2$.}
    \label{fig:Upsilon}
\end{figure}

\begin{figure}[ht!]
    \centering
    \includegraphics[width=0.75\linewidth]{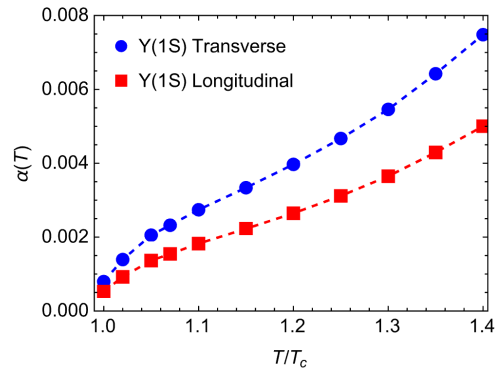}
    \caption{Deviation parameter $\alpha(T)$ of the $\Upsilon(1S)$}
    \label{fig:Upsilon delta}
\end{figure}

To estimate an order of magnitude of the deviation parameter for bottomonium states, we investigated the $\Upsilon(1S)$ as a representative.
The energy ratio of the unpolarized case is shown in Fig.\ref{fig:Upsilon}, but the deviation is hardly observed even at temperatures much higher than $T_c$. This behavior can be understood from OPE structure. Because $d^J(\nu)$ terms in Eq.\eqref{borelOPE} are highly suppressed by the heavy quark mass squared, we can expect that bottomonium states have very small non-trivial 3-momentum dependence compared to the charmonium states. In Ref.\cite{Aarts:2012ka} the energy ratios of $\Upsilon(1S)$ and $\eta_b$ were studied  in a range of $v^2\lesssim 0.04$ using lattice NRQCD, but the authors found no  clear evidence for the non-trivial 3-momentum dependence up to 2.09$T_c$ within uncertainties. 
 In order to extract the deviation parameter of $\Upsilon(1S)$, it requires much effort than charmonium case throughout the numerical analysis because of tiny variation in masses. The deviation parameters for transverse and longitudinal modes are shown in Fig.\ref{fig:Upsilon delta}. As in the $J/\psi$ case, the transverse mode has larger deviation parameter than the longitudinal one. But the magnitude of extracted deviation parameter is very small. For example, even at 1.4$T_c$ and a momentum of 4$\rm{GeV}$($v^2\approx 0.2$), it is expected that the mass shift caused by the 3-momentum is less than 0.01$\%$. Therefore, future lattice simulation may need much higher precision in order to detect meaningful 3-momentum effects on bottomonium states.

\section{Summary and Conclusions}
We investigated the non-trivial 3-momentum effects on the masses of $\eta_c$, $J/\psi$, $\chi_{c0}$, $\chi_{c1}$, and $\Upsilon(1S)$ that are moving in a hot medium using QCD sum rule approach. This study is achieved by including the Wilson coefficients listed in the appendix to the typical QCD sum rule analyses for the quarkonium states at rest. The newly considered Wilson coefficients are responsible for the non-trivial 3-momentum dependence on the ground state mass and explicitly show why the longitudinal and transverse modes in the vector or axialvector channel behave differently in a medium. The non-trivial 3-momentum effects are characterized by the deviation parameter $\alpha(T)$ and our analyses revealed that all quarkonium states experience negative mass shifts as the 3-momentum increases. 
Specifically, we found that the transverse polarization mode in the vector channel has a slightly larger deviation parameter than the longitudinal one, while the axialvector channel has the opposite behavior. 
We also found that the P-wave charmonium states($\chi_{c0}$ and $\chi_{c1}$) experience stronger non-trivial effects than the S-wave ones($\eta_c$ and $J/\psi$). From the OPE structure, we recognized the non-trivial 3-momentum dependence is suppressed by the heavy quark mass squared. Indeed, even if we consider a much higher 3-momentum than used in Ref.\cite{Aarts:2012ka}, the $\Upsilon(1S)$ has a very small deviation parameter compared to charmonium states, e.g. less than a 0.01$\%$ mass shift even at 1.4$T_c$ and at a momentum of 4$\rm{GeV}$.

Meanwhile, it should be noted that our analyses relied on the assumption in which broadening effects are not considered. 
In fact, since the sum rule gives a constraint only on the integral of the spectral function, the mass shift and width broadening have a complementary relationship\cite{Leupold:1997dg}. Therefore, the mass changes computed in this work could be the maximum values that can be expected from the QCD sum rule. But the present work directly observed the non-trivial 3-momentum effects that have not seen yet in the previous lattice simulations, so it will provide a guideline to judge how much uncertainty should be reduced in the future lattice simulations to reveal the genuine 3-momentum effects on the dispersion relation of heavy quarkonium states at finite temperature.

\section*{Acknowledgements}
This work was supported by Samsung Science and Technology Foundation under Project Number SSTFBA1901-04, and by the Korea National Research Foundation under the grant number No.2020R1F1A1075963 and No.2019R1A2C1087107.

\section*{Appendix}
In this appendix, we provide explicit forms of the Wilson coefficients, $d^J(\nu)$, which are responsible for the non-trivial 3-momentum dependence for twist-2 gluon operator. The results are represented by the Whittaker function, $G(b,c,\nu)$, which is defined by
\begin{align}
    G(b,c,\nu)=\frac{1}{\Gamma(c)}\int^\infty_0 dt e^{-t} t^{c-1}(\nu+t)^{-b}.
\end{align}
\begin{widetext}
\begin{align}
    d^{S}&=\frac{\nu}{18 G\left ( \frac{3}{2},\frac{5}{2},\nu \right )}\left\{ 6 G\left ( \frac{1}{2},\frac{7}{2},\nu \right )+8 G\left ( \frac{1}{2},\frac{5}{2},\nu \right )-9 G\left ( -\frac{1}{2},\frac{7}{2},\nu \right ) \right\}, \label{d for S}\\
     d^{P}&=\frac{\nu}{12 G\left( \frac{1}{2},\frac{3}{2},\nu \right )}\left\{ 2 G\left ( \frac{1}{2},\frac{7}{2},\nu \right )+8 G\left ( \frac{1}{2},\frac{5}{2},\nu \right ) - G\left ( -\frac{1}{2},\frac{7}{2},\nu \right )-2 G\left ( -\frac{3}{2},\frac{7}{2},\nu \right ) \right\}, \label{d for P}\\
    d_{L}^{V}&=-\frac{\nu}{6 G\left(\frac{1}{2},\frac{5}{2},\nu\right)}\left\{6 G\left(\frac{1}{2},\frac{7}{2},\nu\right)-6 G\left(-\frac{1}{2},\frac{7}{2},\nu\right) + G\left(-\frac{3}{2},\frac{7}{2},\nu\right)\right\}, \label{d for VL}\\
    d_{T}^{V}&=\frac{\nu}{12 G\left(\frac{1}{2},\frac{5}{2},\nu\right)}\left\{-6 G\left(\frac{1}{2},\frac{7}{2},\nu\right)+8 G\left(\frac{1}{2},\frac{5}{2},\nu\right)+3 G\left(-\frac{1}{2},\frac{7}{2},\nu\right)-2 G\left(-\frac{3}{2},\frac{7}{2},\nu\right)\right\}, \label{d for VT}\\
    d_{L}^{A}&=\frac{\nu}{3 G\left ( \frac{3}{2},\frac{5}{2},\nu \right )}\left\{  G\left ( \frac{1}{2},\frac{7}{2},\nu \right )-2 G\left ( -\frac{1}{2},\frac{7}{2},\nu \right )\right\}, \label{d for AL}\\
    d_{T}^{A}&=\frac{\nu}{12 G\left ( \frac{3}{2},\frac{5}{2},\nu \right )}\left\{ -2 G\left ( \frac{1}{2},\frac{7}{2},\nu \right )+8 G\left ( \frac{1}{2},\frac{5}{2},\nu\right )-5 G\left (-\frac{1}{2},\frac{7}{2},\nu \right) \right\}. \label{d for AT}
\end{align}
\end{widetext}

\bibliographystyle{apsrev4-1}
\bibliography{refs.bib}

\end{document}